# Density and well width dependences of the effective mass of two-dimensional holes in (100) GaAs quantum wells measured by cyclotron resonance at microwave frequencies

H. Zhu[1], K. Lai[2] *, D. C. Tsui[2], S. P. Bayrakci[1], N. P. Ong[1], M. Manfra[3], L. Pfeiffer[3], K. West[3]

1. *Department of Physics, Princeton University, Princeton, New Jersey 08544, USA*
2. *Department of Electrical Engineering, Princeton University, Princeton, New Jersey 08544, USA*
3. *Bell Labs, Lucent Technologies, Murray Hill, New Jersey 07974, USA*

## Abstract

Cyclotron resonance at microwave frequencies is used to measure the band mass ($m_b$) of the two-dimensional holes (2DHs) in carbon-doped (100) GaAs/Al$_x$Ga$_{1-x}$As heterostructures. The measured $m_b$ shows strong dependences on both the 2DH density (p) and the GaAs quantum well width (W). For a fixed W, in the density range ($0.4 \times 10^{11}$ to $1.1 \times 10^{11}$ cm$^{-2}$) studied here, $m_b$ increases with p, consistent with previous studies of the 2DHs on the (311)A surface. For a fixed p = $1.1 \times 10^{11}$ cm$^{-2}$, $m_b$ increases from $0.22 m_e$ at W = 10nm to $0.50 m_e$ at W = 30nm, and saturates around $0.51 m_e$ for W > 30nm.

* Present address: Dept of Applied Physics, Stanford University, Stanford, CA 94305.

For a two-dimensional (2D) system confined in GaAs/Al$_x$Ga$_{1-x}$As heterostructures, the structure of the valence band is considerably more complicated than that of the conduction band, where a constant electron effective mass well captures the band structure. Due to the 2D quantum confinement, the heavy hole sub-band admixes with the light hole sub-band, giving rise to a highly non-parabolic valence subband structure [1-3]. As a result, the hole effective mass (m$_b$) is a function of many sample-dependent parameters, including the confining potential (the barrier height, the quantum well width, the doping configuration, etc), the 2D hole density (p), and the Miller index of the grown surface. A full theoretical description of the band structure of the 2D holes is thus a tedious and formidable task [4]. On the other hand, m$_b$ in a GaAs 2D hole system (2DHS) is a very important sample parameter for device application, quantum transport, and optical study, and there have been a number of experimental investigations addressing this issue. However, the earlier cyclotron resonance (CR) experiments were carried out in the high density limit with $p > 1 \times 10^{11}/cm^2$ at far infrared frequencies, probing large Landau level separations in the meV energy range [Ref. 5-7]. Only recently advances in heterostructure crystal growth technology have made it possible to materialize 2DHS with density $p \leq 1 \times 10^{11}/cm^2$ and mobility sufficiently high to allow low excitation energy microwave CR in the GHz frequency range [Ref. 8,9].

In physics research of the 2DHS, where high mobility is essential, much work has been done in the silicon-doped samples grown on (311)A GaAs substrates [10]. While the mobility achieved in these 2DHSs is quite impressive, several limitations related to the (311)A surface have been recognized, e.g., the reduced symmetry and the surface corrugation [11]. Recent progress in two types of (100)-oriented GaAs 2DHSs effectively removes these constraints. First, the p-type heterojunction insulated-gate field-effect transistors (HIGFETs) on (100)-GaAs substrates are now routinely fabricated to study the 2D metal-insulator transition (MIT) [12, 13]. Second, instead of silicon, carbon is now used for modulation doping in the molecular-beam epitaxy (MBE) growth to produce (100) holes with mobility over $10^6$ cm$^2$/Vs [14]. In both cases, the m$_b$ of (100) 2DHS is needed for probing into the 2D physics. In particular, the strength of the hole-hole interaction is determined by the ratio of the Coulomb energy to the Fermi energy, with the latter inversely proportional to m$_b$.

In this letter, we report an experimental determination of m$_b$, using the cyclotron resonance (CR) technique, of the 2DHSs confined in (100) carbon-doped GaAs/Al$_x$Ga$_{1-x}$As heterostructures. The measured m$_b$ shows strong dependences on both the 2DH density ($p$) and the GaAs quantum well width ($W$). For a fixed $W$, $m_b$ increases with $p$ in the range from $0.36 \times 10^{11}$ to $1.1 \times 10^{11}$ cm$^{-2}$, consistent with previous studies on the (311)A 2DHSs. For a fixed $p = 1.1 \times 10^{11}$ cm$^{-2}$, $m_b$ increases from $0.22 m_e$ at $W = 10$nm to $0.50\ m_e$ at $W = 30$nm, and saturates around $0.51 m_e$ for $W > 30$nm.

We have investigated a total of ten samples in this study, all carbon-doped (100) GaAs/Al$_x$Ga$_{1-x}$As heterostructures grown by MBE. Except for sample 3, the (asymmetric) δ-doping is from the front side only. Important parameters of the individual samples are listed in Table 1, including the Al mole fraction x, the density p, the quantum well width W and the mobility μ measured at 0.3K, together with the band mass m$_b$. For most samples, the DC transport data can be found in Ref. [14].

The same cyclotron resonance technique, previously employed for studying the (311) A p-GaAs samples [10], was used to measure $m_b$. During a single measurement, the microwave frequency (f) was fixed and the magnetic (B) field was swept. Near the resonance, the 2DHS sample, thinned to 100 ~150 μm and glued to a sapphire substrate, absorbed the microwave radiation power and its temperature rose above 4.2K. This temperature difference was detected by an adjacent bolometer glued to the same substrate. In the linear response regime, where our measurements were carried out, the CR amplitude was proportional to the real part of the AC complex conductivity of the 2DHS, which showed a resonance under the condition of $\omega_c = 2\pi f = 2\pi e B_{CR}/m_b$, where $B_{CR}$ was the resonance magnetic field. Further experimental details can be found in Ref. [15].

The two insets in Fig. 1 show the magnetic field dependence of the microwave absorption signal obtained at a fixed frequency of 40 GHz for three different samples, as well as for the same sample at three different frequencies. All curves display a well-defined resonance peak and the line shape of these curves can be fitted by the Drude model [16], shown as the solid lines. In Fig. 1, f is plotted as a function of the peak magnetic field $B_{CR}$ for sample 2. For f > 35 GHz, a linear f vs $B_{CR}$ (dashed line) is observed and a mass of 0.33 $m_e$ is deduced from the slope. For f < 20 GHz, however, the data deviate from the linear dependence and the resonance splits into two branches. This behavior is due to coupling of the edge magnetoplasmon (EMP) mode to the cyclotron resonance mode [17]. When this coupling occurs near the EMP frequency $\omega_0$ at B = 0 T, the resonant frequency is modified to $\omega_\pm = \pm\omega_c/2 + [(\omega_c/2)^2 + \omega_0^2]^{1/2}$. The thick solid lines in Fig. 1 show the fits to both the low frequency and high frequency parts. The result, $m_b$ = 0.33 $m_e$, agrees with that from the linear fit for high frequencies. Fitting to the Drude model and the EMP mode has been discussed in our previous paper [9] and will not be repeated here.

Fig. 2 summarizes the results of $m_b$ in our samples. The dependences of $m_b$ on both W and p are readily seen from the plot. For fixed W's at 10nm, 15nm, or 30nm, in the density range ($0.36 \times 10^{11}$ to $1.1 \times 10^{11}$ cm$^{-2}$) studied here, $m_b$ increases with p, consistent with previous studies from the 2DHSs on the (311)A surface [9]. For a fixed $p = 1.1 \times 10^{11}$ cm$^{-2}$, $m_b$ of the four single-side doped samples (Sample No. 1, 2, 4, and 5) increases monotonically as a function of W and saturates at 0.51 $m_e$ for W > 30nm. A slightly larger mass $m_b = 0.54\ m_e$, however, is obtained for sample 3 with W=20 nm, which is probably due to the more symmetric hole wavefunction in this double-side doped QW sample. Our data clearly manifest the complicated valence band structure in (100) GaAs quantum wells.

The density and well width dependences of $m_b$ can be qualitatively understood by the non-parabolic nature of the 2D valence bands [1]. In a quantum well, the four-fold degeneracy of the J=3/2 valence band at the center of the Brillouin zone is removed and the lowest energy hole subband is the $J_z=\pm 3/2$ heavy hole subband (HH1), which has a small in-plane effective mass. The energy gap separating it from the $J_z=\pm 1/2$ light hole subband (LH1) depends strongly on the quantum confinement. Away from the zone center, the HH and LH bands mix giving rise to level anti-crossing and highly non-parabolic, anisotropic in-plane dispersion. At a fixed W, the non-parabolic effect is small when density is low and the Fermi contour is nearly a circle. The $m_b$ is heavier at higher densities because the non-parabolic effect becomes important. In fact, it has been shown that the Fermi contour here is significantly distorted [18]. For samples with different W's, the confinement is stronger in smaller W samples, giving rise to a larger band splitting. As illustrated in the insets of Fig. 2, for narrow well samples with W = 10nm, the splitting is so large that the measured $m_b$ = 0.21 $m_e$ for p = 0.44x10$^{11}$/cm$^2$ and 0.22$m_e$ for p = 1.1x10$^{11}$/cm$^2$, practically independent of p. In fact, in our previous experiments in (311)A 2DHSs, a mass of 0.19$m_e$ is also observed for the 10nm narrow well sample [10]. For wide wells with W > 30nm, on the other hand, the splitting at the zone center is small. As a result, even in our low density sample with p = 0.36×10$^{11}$ cm$^{-2}$, the Fermi level $E_F$ may already be at an energy sufficiently away from the HH1 and LH1 level anticrossing that $m_b$ only weakly depends on p. Interestingly, it is suggestive from our limited data points that for W = 15 ~ 20nm, the strong density dependence of $m_b$ is due to the subband non-parabolicity in the level anti-crossing region of the in-plane dispersion. Experiments are underway to measure $m_b$ for various p's at a fixed W = 20nm. We nevertheless want to emphasize that, the real band structure of the 2DHS has a much more complicated shape and our oversimplified picture can explain the data only qualitatively.

Finally, two remarks on the data are in order. First, in Fig. 3(a) we compare the $m_b$ vs p relation for the (311)A [10] and the (100) p-GaAs 2DHS at the same quantum well width W = 30nm. It appears that the density dependence of $m_b$ is weaker for 2DHS in the (100) orientation than that in (311)A. Second, our data help to shed some light on the puzzling transport data previously reported by Manfra et al. [15]. In Fig. 3(b), we plot the transport mobility ($\mu$) measured at T = 0.3K and the DC relaxation time ($\tau_{DC} = \mu m_b/e$) as a function of the well width W at p ~ 1×10$^{11}$ cm$^{-2}$. It is clearly seen that $\mu$ in these samples shows a sharp peak at W = 15nm, which is in contrast to the case in GaAs 2D electron system with monotonic increasing $\mu$ vs W. It was conjectured that this anomalously high $\mu$ at W

= 15nm may not be due to a large $\tau_{DC}$, but instead the change of other material parameters like $m_b$. Indeed, we observe in Fig. 3(b) that, because of the smaller $m_b$ at W=15nm than that at 30nm, $\tau_{DC}$ is actually shorter at W=15nm. However, even after taking into account the effect of $m_b$, $\tau_{DC}$ = 0.21ns at W =30nm is still longer than that in wider wells.

To summarize, we have experimentally determined the band mass of 2DHSs in carbon-doped p-type (100) GaAs/Al$_x$Ga$_{1-x}$As heterostructures. The measured $m_b$ shows strong dependences on both the 2D hole density p and the well width W. The fact that $m_b$ increases with either p or W is qualitatively consistent with the nature of non-parabolic 2D valence bands.

The authors would like to thank Wei Pan, Minhyea Lee, and Lu Li for helpful discussions and experimental assistance. The work at Princeton was supported by the NSF MRSEC grant DMR 0213706.

Table 1. Sample parameters, including the Al mole fraction x, the quantum well width W, the 2DH density p, the mobility $\mu$ at 0.3K, and the measured cyclotron mass $m_b$.

| Sample | x (%) | W (nm) | p ($10^{11}cm^{-2}$) | $\mu$ ($10^5 cm^2/Vs$) | $m_b$ ($m_e$) |
|---|---|---|---|---|---|
| **1**(8-29-03.2) | 32 | 10 | 1.1 | 5.9 | 0.22 |
| **2**(9-8-03.1) | 32 | 15 | 1.1 | 10.0 | 0.33 |
| **3**(1-11-05.1) | 16 | 20 | 1.1 | 20 | 0.54 |
| **4**(9-8-03.2) | 32 | 30 | 1.1 | 7.3 | 0.50 |
| **5**(8-21-03.3) | 32 | Single Interface | 1.1 | 5.2 | 0.51 |
| **6**(10-30-03.2) | 32 | 50 | 0.9 | 5.7 | 0.48 |
| **7**(11-17-04.2) | 16 | 10 | 0.44 | 8 | 0.21 |
| **8**(12-24-03.2) | 16 | 15 | 0.36 | 6.6 | 0.24 |
| **9**(12-24-03.1) | 16 | 30 | 0.36 | 6.6 | 0.48 |
| **10**(12-16-04.1) | 16 | 20 | 0.56 | 22 | 0.39 |

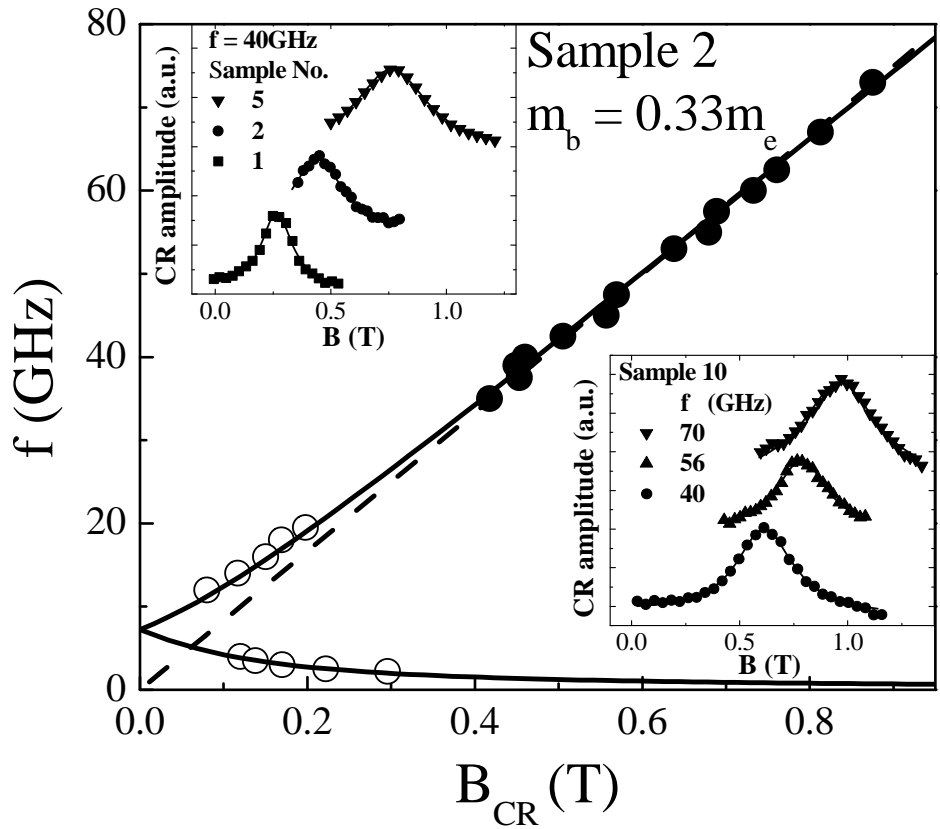

Fig. 1 Microwave frequency f as a function of the resonance magnetic field $B_{CR}$. The dashed line shows the linear fit to the high-f data through zero, which gives a band mass $m_b = 0.33 m_e$. The solid line is the fit to all data points (filled and empty circles) using the EMP formula in the text. The inset in the top-left shows the CR signal vs B-field for three different samples at f = 40GHz, and the bottom-right inset for sample 10 at three different frequencies.

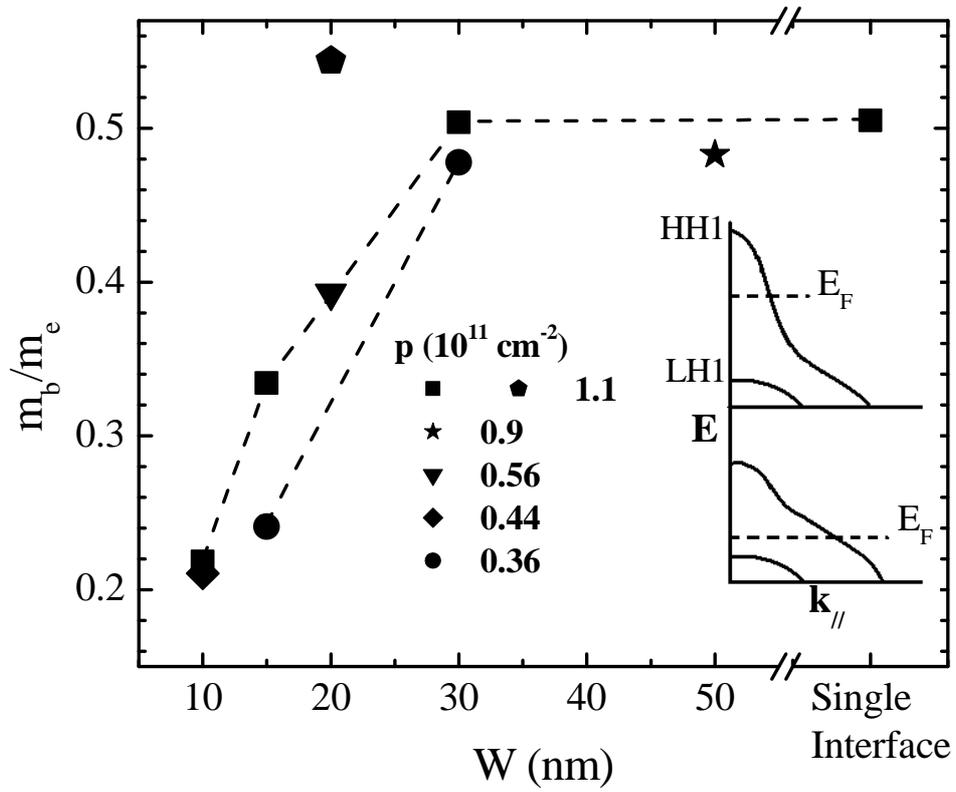

Fig. 2 $m_b$ as a function of W at various 2DH densities in this study (single interface sample has W=1000nm). The dashed lines are guides to the eyes. The two schematics show the possible valence band structures for (upper) narrow quantum wells and (lower) wide quantum wells. $E_F$ in both schematics shows the position of the Fermi level.

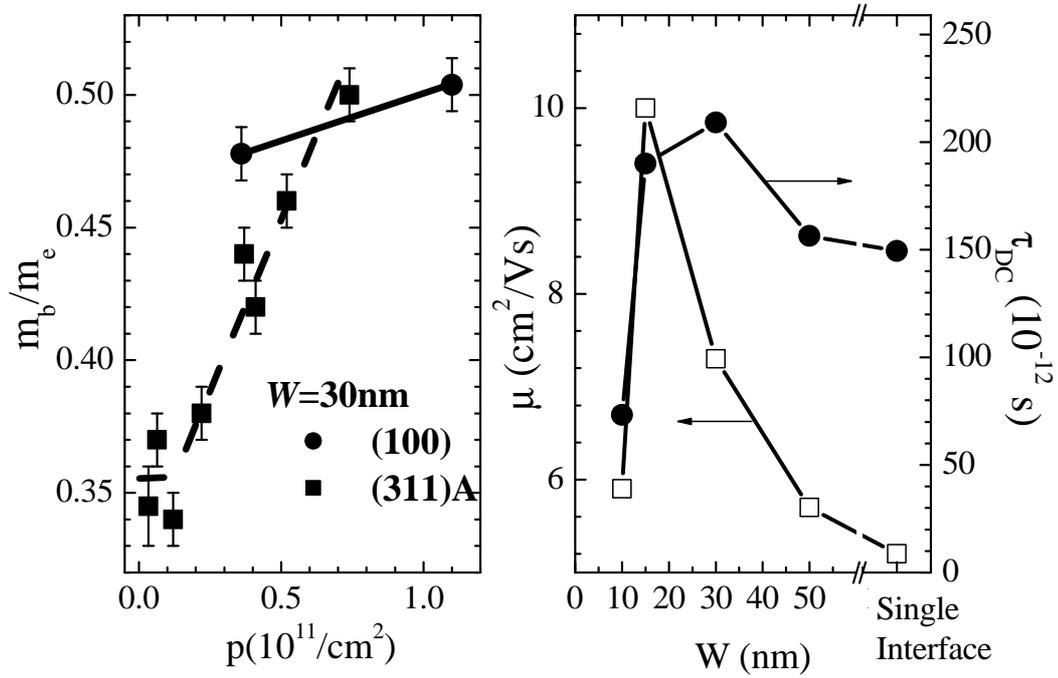

Fig. 3 (a) Comparison of $m_b$ vs p between samples grown on (100) and (311)A substrates. The well width is the same, 30nm, for both cases. The lines are guides to the eyes. (b) Transport relaxation time $\tau_{DC}$ and mobility $\mu$ vs the well width W at p ~ $1\times10^{11}$cm$^{-2}$ (single interface sample has W=1000nm). The mobility data were taken at T = 0.3K.